# Самоорганизация перколяции токовых каналов в среде с тепловой нелинейностью проводимости: сопоставление экспериментальных данных и компьютерного моделирования в рамках каскадной модели гигантских флуктуаций проводимости полимерных композитов.

### Введение

Полимеры и в особенности полимер-композиты, как известно, представляют собой классические сильно-неоднородные материалы, имеющие многомасштабную пространственную структуру, вообще говоря, фрактального типа [1]. В базовом полимере основной ( и практически единственной) «точной» характеристикой является собственно молекула мономера, образующего макромолекулы полимера, а практически все остальные параметры, включая длины и массы макромолекул, степень кристалличности, наличие полимер-изомерных структур, свободный объем, плотность, проводимость и др. - зависят как от способа получения полимера, так и от многих других параметров подготовки изделия или образца. В силу отмеченной выше нано- и микро-неоднородности в композитных материалах как следствие возникают микро-неоднородности полей и токов и, соответственно, относительно легко (по сравнению с однородной средой) могут проявляться нелинейные эффекты, что вызывает неослабевающий интерес к исследованию полимерных образцов [2] и практическим приложениям композитных материалов [3]. В частности, в работе [3] была экспериментально исследована и предложена физическая гипотеза функционирования полимер-композитного термистора с положительным ТКС. В экспериментах [3] проводящими элементами (в матрице полимера полиэтилен и/или эпоксидные смолы) были металлические частицы с концентрацией выше порога перколяции. Вследствие электроразогрева композита выше порога перколяции [3], приблизительно до точки стеклования полимера, наблюдался переход из проводящего (перколяционного) состояния высокой проводимости (СВП) в состояние низкой проводимости (СНП), т.е. фактически диэлектрика, с перепадом проводимости на 12 порядков величины [3], что собственно и использовалось на практике для ограничения тока. Физическая модель этого явления на качественном уровне обсуждалась в [3], причем среди возможных физических процессов переключения сопротивления указывалось то, что при нагреве и расширении композита полимер расширяется сильнее и проводящие частицы, которые ранее соприкасались, разделяются промежутками. Другая модель, описанная там же, связана с процессом возникновения областей кристаллизации полимера, способствующим неоднородному, более плотному распределению частиц и резкому прохождению порога перколяции в этой температурной точке.

В настоящей работе приводятся результаты численного счета в рамках модели тепловой кондуктивной неустойчивости созданной в развитие гипотезы [4,5] посредством прямого численного моделирования с привлечением уравнения теплопроводности и проводятся сопоставления с экспериментальными известными в литературе результатами [2,6], а также полученными авторами ранее[7,8] по переключениям сопротивления на те же 12 порядков (что и в [3] только с отрицательным ТКС). Эксперименты выполнялись на серии образцов композита на основе ПВХ с проводящими элементами в виде фрагментов молекул ПолиАцетилена (ПАц) с отрицательным ТКС. Наиболее удивительным результатом экспериментальных исследований скачков проводимости в сополимере ПВХ-ПАц следует считать отчетливо наблюдаемые спонтанные переключения проводимости на много порядков величины в некоторых случаях — более 11 порядков величины[6,7]. Осциллограмма тока на образце для одного из таких спонтанных переключений приведена на рис.1. Можно обратить внимание на то обстоятельство, что как до появления спонтанного скачка тока (перехода образца в СВП), так и после возврата в СНП отчетливо наблюдаются многочисленные более

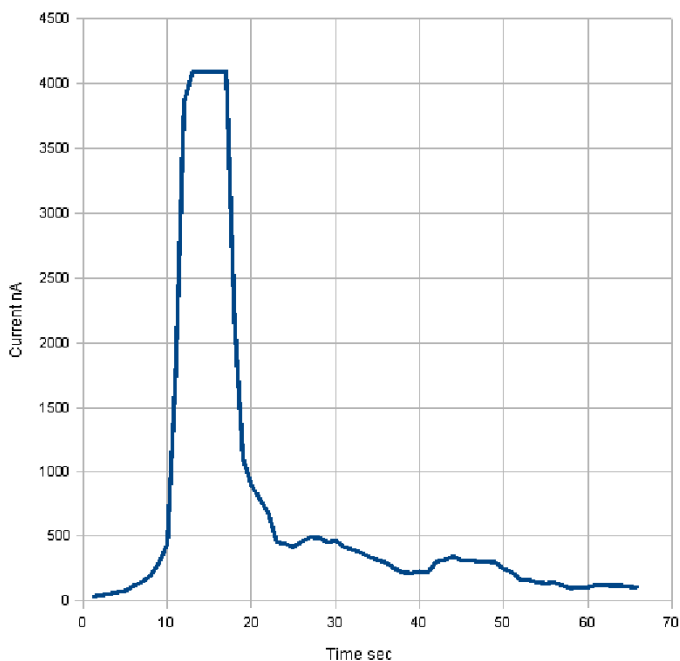

Рис.1 Осциллограмма спонтанного переключения проводимости и возврата в исходное состояние для образца сополимера ПВХ-ПАц [4] . Ограничение амплитуды импульса сверху связано с ограничением, вносимым предельной шкалой цифрового осциллографа.

Частота повторения спонтанных выбросов тока зависит от многих параметров и первоначально переходы существенно затрудняли процесс измерения сопротивления. Мы обратили внимание на реальность этих переходов, когда стали регистрировать длинные компьютерные записи токовых сигналов для выявления процессов релаксации.

 На рис.2 приведена измеренная область концентраций и величина максимального скачка

Обобщенные по серии образцов измерения амплитуды и вероятности спонтанных переходов в зависимости от концентрации фрагментов ПСС в ПВХ, т.е. времени дегидрохлорирования приведены на рис. 2.

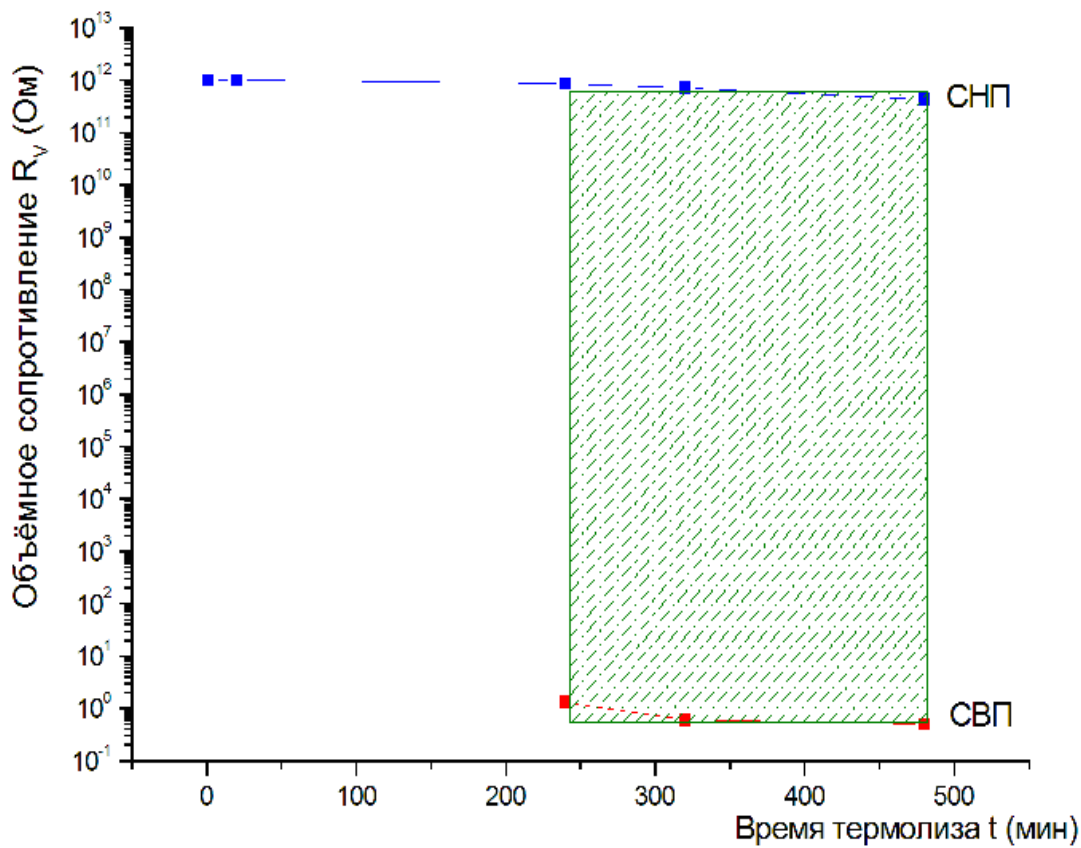

Рис.2. Область концентраций ПСС (времени дегидрохлорирования ), в которых происходят спонтанные переходы СНП-СВП, а по оси абсцисс максимальная амплитуда изменения сопротивления (в логарифмическом масштабе).

В отличие от работы [3] компьютерное моделирования процесса самоорганизации перколяционного перехода в полимерных композитах на основе ПВХ осуществлялось ниже порога перколяции в исходном состоянии. Прежде всего напомним о том, что представляет собой перколяция в тонкой пленке сильно неоднородного композитного материала.
       Для этого воспользуемся классической работой, в которой подробно изложены приложения теории перколяции и анализа протекания тока в сильно неоднородных средах [1], в том числе вблизи порога перколяции. В частности, показано, что в композитной среде вблизи порога перколяции образован Бесконечный Кластер (БК), т.е. превышен порог перколяции, а для пленочных образцов (фактически двумерных систем) порог протекания и собственно БК может быть связан с образованием одного (вблизи порога перколяции) или нескольких тонких токоведущих каналов, что, кстати, неоднократно наблюдалось в экспериментах [2]. И этот вывод вытекает из простейшего, чисто умозрительного, рассуждения о том, что, если из трехмерного БК вырезать тонкие «пленочные» образцы, то вблизи порога перколяции в таком образце может оказаться один или несколько токоведущих каналов. Нас интересует именно эта ситуация, когда концентрация энергии происходит в одном узком, порядка десятков и менее микрон, токоведущем канале, поскольку, как показывают оценки [4,5], в этом случае могут возникать нелинейные эффекты и наблюдаться процесс самоорганизации перколяции с резким перепадом сопротивления ( в наших экспериментах до 12 порядков величины [6]) вблизи порога перколяции. Ранее для композитных материалов, содержащих проводящие включения и изолирующие промежутки, предложена универсальная каскадная модель [4] развития неустойчивости за счет аномально высоких флуктуаций поля вблизи порога перколяции, где в полимерах и в полимер-композитных материалах может реализоваться квази-лавинный «мягкий» обратимый пробой.

Можно сказать, что аналогичный мягкий, обратимый пробой наблюдался и в работе [3] - полимерный термистор-  где при нагревании  вблизи и выше точки стеклования полимера наблюдался переход  из проводящего состояния в состояние фактически диэлектрика с перепадом проводимости на 12 порядков величины. В настоящей работе развита компьютерная модель, описывающая противоположный эффект — самоорганизация перколяционного перехода при слабых, «измерительных» токах порядка наноампер, также опирающаяся на  экспериментальные наблюдения с переключением проводимости более чем на 11 порядков величины, но в обратную сторону, т.е. переход из состояния  низкой проводимости (СНП) - практически изолятора [6,8] в состояние высокой проводимости (СВП), близкое к  уровню проводимости металла.  Описаны необычные свойства нанокомпозита на основе сополимера ПВХ-ПАц, причем фрагменты молекул ПАц  встроены в макромолекулы ПВХ, который  представляет собой некий аналог композита с углеродными нанотрубками [9], где в отличие от работы [3] проводящими элементами являются фрагменты двойных сопряженных связей. Получение нанокомпозита ПВХ-ПАц реализуется в подробно описанном  в литературе процессе термолизации (отделения молекул H-Cl от макромолекулы ПВХ в растворе с образованием цепочек ПСС). Среди наиболее необычных свойств исследуемого сополимера был выявлен ряд  гигантских скачков проводимости, вызываемых как изменением приложенного напряжения (в процессе измерения вольтамперных характеристик), так и самопроизвольными спонтанными переходами при неизменных внешних условиях (отметим, что низковольтные спонтанные переходы  в композитах наблюдались также в [6]).  В наших экспериментах переход происходит между состояниями относительно быстро в то время, как сохранение состояний заданной проводимости  как  на уровне СВП, так и в исходном СНП образец может находиться  в течение промежутка времени не сопоставимо большего, чем собственно время перехода между состояниями. Ключевым моментом модели, предложенной в [5], является  механизм задержки состояний СВП и возврата в СНП вследствие конечного времени блуждания освободившейся молекулы H-Cl  в локальном «прилежащем» участке свободного объема полимерной матрицы. Именно эта задержка может обусловливать осцилляции температуры, гистерезис и непредсказуемость вольт-амперной характеристики нанокомпозитов  с «зависанием» на неопределенное время образцов в одном из фиксированных состояний проводимости.

2. Модель численного моделирования.

Собственно решалась линеаризованная задача самовоздействия токов (через механизм разогрева макромолекул ПВХ и локального дегидрохлорирования (ДГХ) с соответствующим уменьшением сопротивления) на один или несколько изолирующих промежутков в токоведущем канале с размерами менее 10 мкм  Для этого последовательные итерации значений тока, нагрева и сопротивлений изолирующих промежутков находились посредством многократно решаемого неоднородного уравнения теплопроводности с малыми приращениями указанных параметров:

$U_t = a^2 \Delta U + f(x,t)$ , (1)

где U – температура, $U_t$ частная производная по температуре, $\Delta$ -оператор Лапласа, f(x,t)- источник Джоулева тепла за счет электро разогрева, $a^2$- коэффициент температуропроводности полимера, определяемый через коэффициент теплопроводности $\chi$ выражением :

$a^2 = \dfrac{\chi}{\rho \cdot C}$ , где $\rho$ — плотность, а C — изобарная теплоемкость полимера.

Для задачи плоского образца нанокомпозита между двумя плоскими проводящими электродами трехмерное уравнение за счет разделения переменных сводится к двумерному (температура электродов постоянна).

В цилиндрическом токоведущем канале уравнение (1) преобразуется к известному виду:

$$\frac{a^2}{r} \cdot \frac{\partial}{\partial r} r \frac{\partial T}{\partial r} + F(r,t) = \frac{\partial T}{\partial t} \qquad (2),$$

причем граничные условия по r берутся стандартными для неограниченного тела - полимер считается однородным, а расстояние до соседних проводящих элементов пренебрежимо велико. Однородное уравнение, соответствующее неоднородному уравнению (1) имеет известное фундаментальное решение задачи Коши, откуда с помощью формулы Пуассона находим Функцию Грина уравнения (2), через которую выражается решение неоднородного уравнения (при однородных граничных условиях) в стандартном виде:

$$U(r,t) = \int_0^t \int_{-\infty}^{\infty} f(\zeta,\tau) * G(r,\zeta,t-\tau) d\tau d\zeta \qquad (3),$$

далее интеграл (3) можно значительно упростить, допуская, что зависимость f(x,t) медленно зависит от t и далее можно учесть эту зависимость последовательными итерациями по времени и соответственно по изменениям температуры. Далее получим из (3):

$$U(r,t) = \frac{1}{2a} \cdot \int_0^{\infty} \int_0^t \left( \frac{\xi}{(t-\theta)} \exp\left(\frac{-(r^2+\xi^2)}{4a \cdot (t-\theta)}\right) f(\xi) I_0\left(\frac{r \cdot \xi}{2a \cdot (t-\theta)}\right) \cdot d\xi d\theta \right) \qquad (4),$$

численно можно рассчитывать как распределение температурного поля по пространству r, так и вычислять оценку температуры непосредственно на оси канала, а упростив (4) еще больше, что при наличии идентичных и достаточно плавных пространственных распределений температуры по r, можно получить уравнение, которое может оказаться вполне достаточным для решения поставленной задачи. Выписанное решение (4), это только начало решения нелинейной задачи кондуктивной неустойчивости, связанной с зависимостью сопротивлений изолирующих промежутков (ИП), составляющих по нашим экспериментальным данным суммарно по образцу от $10^{-8}$ до $10^{-11}$ Ом. Неоднородный член уравнения (1) для i-го изолирующего элемента в токоведущем канале имеет вид:

$$f_i(r,t) = \frac{I^2(t) R_i(t)}{V \rho C} \qquad (5),$$

причем ток I(t) в цепи определяется законом Ома для полной цепи:

$$I(t) = \frac{U_0}{\sum_{i=1}^{max} R_i(t) + R_s + R_b} \qquad (6),$$

где сумма $R_i$ -изолирующие элементы в токовом канале, $R_i$ -остаточное (не зависящее от времени) сопротивление образца и $R_b$ - балластное сопротивление измерительной цепи, в которое входит внутреннее сопротивление терраомметра и при необходимости дополнительное балластное сопротивление, используемое для согласования с цифровым осциллографом и автоматизированной считывающей аппаратуры. В ходе экспериментальных измерений обычно сопротивление образца в СНП превышает все остальные на несколько порядков, и наоборот, при переключении в СВП ток практически полностью определяется баластным сопротивлением и фактически становится постоянным. Это наводит на мысль, что линеаризация системы параметризацией системы по времени , т. е. собственно зависимость сопротивления элемента от температуры существенна только при высоких температурах, когда согласно модели происходит реакция дегидрохлорирования и сопротивление $R_i$ – зависит от температуры только при высоких температурах, оставаясь неизменным в широком диапазоне низких температур. Это подсказывает метод-алгоритм счета: для фиксированных $R_i$ - для тока (обычно в диапазоне наноАмпер) решая неоднородное уравнение теплопроводности, находим приращение температуры ΔU , затем пересчитываем новые значения полного сопротивления образца, получаем новое значение тока I(t+Δt) и с новым значением функции источника вновь решаем уравнение (1). Аналитически обосновать выбранный метод расчета достаточно сложно, однако «инертность» уравнения теплопроводности, определяемая характерным временем отклика $\tau = \dfrac{1}{a^2 \cdot r_o^2}$, позволяет выбрать шаг существенно меньший, чем указанное характерное время, и можно сколь угодно близко приближаться к истинному решению нелинейной системы. Хотя приведенное обоснование не является строгим, оно представляется достаточно разумным.

3. Результаты моделирования.

Для простоты и понимания процессов, происходящих в нелинейной системе, моделирование проводилось первоначально для одного изолирующего промежутка $R_1$, при этом результаты расчета приведены на рис.3. Если взять несколько элементов $R_i$, в частности при i=3 при токе порядка одного наноАмпера все элементы испытывают описанные выше нелинейные непериодические осцилляции.

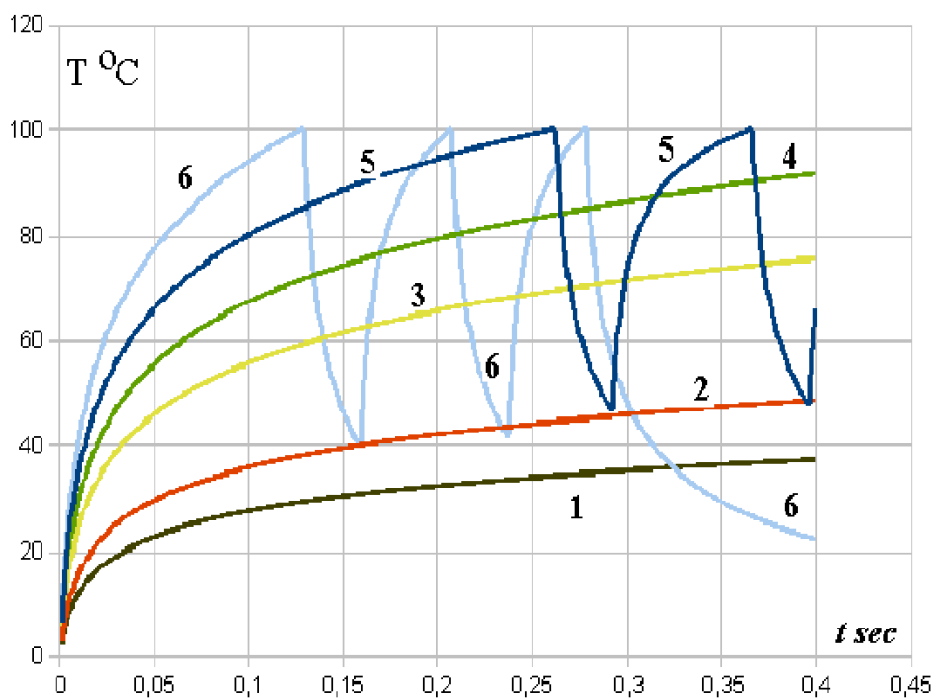

Рис.3. Вариация температур на одном элементе сопротивления — одном изолирующем промежутке, наблюдается несколько характерных «мод» поведения, кривые 1, 2, 3, кривые, никогда не достигающие температур дегидрохлорирования $T_d$, фактически выходят на стационарное решение на уравнение теплопроводности, кривые 5, 6, осциллирующие за счет достижения $T_d$ и переключения за счет выделения H-Cl в свободный объем полимера, после чего следует этап охлаждения, когда $R_i$ – в СВП состоянии имеет низкое сопротивление и согласно (5) тепловыделение резко уменьшается — тепло рассеивается на $R_s$ и $R_b$, когда (двойная сопряженная связь разрушается, в частности присоединяя молекулу HCl) происходит обратный переход в СНП и снова начинается разогрев. В модельных расчетах размер (диаметр) канала 8 мкм, размер элемента 3 мкм, сопротивление в СНП $10^9$ Ом с учетом полного сопротивления цепи -максимальный ток в цепи- несколько нано Ампер.

Таким образом, полная нелинейная система уравнений включает уравнение теплопроводности (1), с неоднородным источником (5) и уравнение - Закона Ома (6) для полной измерительной цепи. Все уравнения по отдельности достаточно просто решаются численным методом, в частности, с использованием специальной хорошо известной расчетной программы "Maxima", которая и использовалась в основном в данных расчетах.

Собственно «разорвать» нелинейную систему, как можно предположить с минимальными потерями точности, удалось учитывая, что уравнение теплопроводности достаточно «иннерциально» по времени с приведенным выше характерным временем τ, поэтому выбирая шаг по времени в рамках малой доли τ, можно получить «устойчивые» решения практически независимые от шага Δτ. Таким образом, анализ поведения одного изолирующего промежутка в зависимости от приложенного напряжения в предположении существования одного канала 10 мкм, показывает, что температурная зависимость элементов по времени может быть плавной, стремящейся с пределу, определяемому решением стационарного уравнения. При более высокой теплоотдаче температура в изолирующих промежутках достигает температуры ДГХ, при этом вместо $R_i$ изолирующего промежутка возникает проводящая цепочка ПСС. Соответственно источник $I^2 R_i$ резко уменьшается, причем, если сопротивление падает согласно гипотезе на 10-11 порядков величины, то изменения тока существенно меньше, поскольку определяются суммой всех сопротивлений, включая балластное сопротивление, остаточное внутреннее сопротивление (6). Таким образом, участок ПВХ молекулы, превращенный в проводящую ПСС цепочку, начинает остывать и повышается вероятность обратного ДГХ преобразования, т.е. присоединения молекулы H-Cl, находящейся в близлежащем элементе свободного объема с переходом в состояния изолятора. Таким образом, наблюдаемые экспериментально переключения по развиваемой модели связаны с микропереключениями изолирующих элементов композита.

При увеличении числа изолирующих промежутков в одном канале поведение каждого из элементов остается приблизительно таким же, как и на рис.3., в то время как ток в цепи изменяется существенно меньше, за исключением ситуации, когда нагрев и переключение сопротивления $R_i$ происходят синхронно.

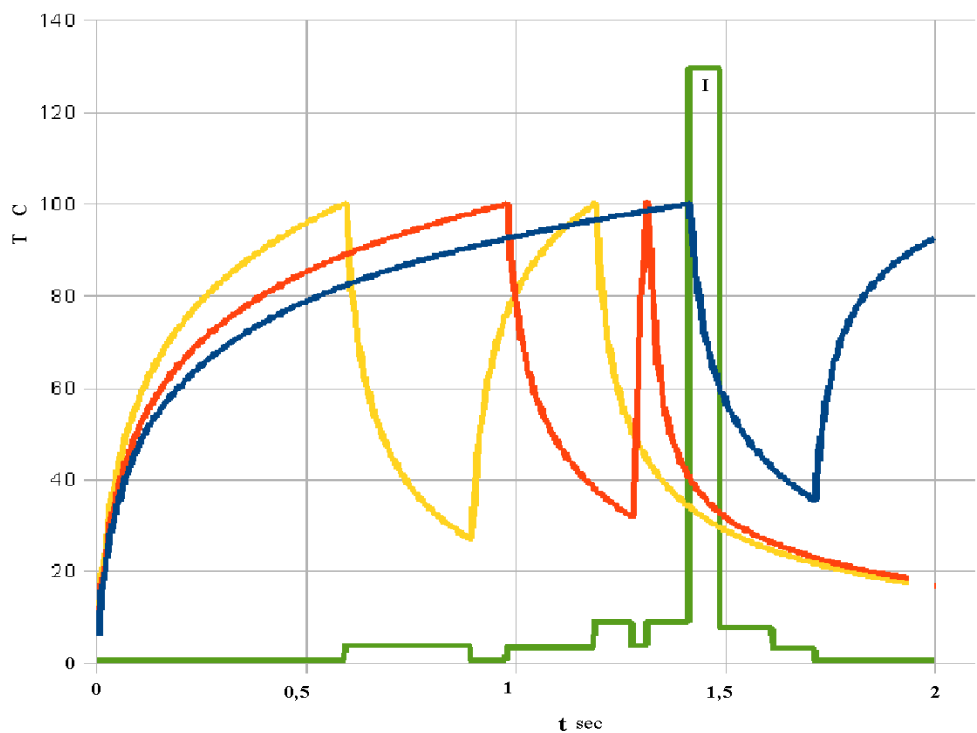

Рис.4. Синхронизация процесса злектроразогрева 3х изолирующих промежутков: температура каждого их трех элементов (в порядке возрастания максимального значения $R_i$ ) показана соответственно, желтым, красным и синим. Зеленым цветом показано «наблюдаемое» гигантское переключение сопротивления на 10 порядков, фактически спонтанно возникающим в некоторый момент времени, аналог экспериментального, показанном на рис.1.

Кроме тривиального и неинтересного случая, когда все параметры изолирующих промежутков одинаковы, возможна синхронизация нескольких разно периодных квазипериодических процессов. Так на рис. 4. приведена картина, возникающая при синхронизации процессов в трех существенно различных по сопротивлению (толщине) изолирующих промежутков.

В соответствие с проведенными расчетами «спонтанность», наблюдавшаяся в наших экспериментах, а также в работе [9] согласно развиваемой гипотезе и выполненным расчетам для композитов на базе ПВХ или точнее момент возникновения переднего фронта спонтанного импульса связан с взаимной синхронизацией циклов ИП, причем температурные циклы последовательно расположенных элементов не независимы между собой, в частности, как видно из рис. 4. после открытия двух ИП ток и выделение тепла на третьем элементе возрастает и возникает некое подобие самосинхронизации. Этим объясняется передний фронт спонтанного импульса, при этом ток максимален и определяется только остаточным и балластным сопротивлением, что, как правило, в сотни и более раз ниже полного сопротивления. При этом все элементы охлаждаются и повышается вероятность перехода в ИП. Однако первый ИП прогревается максимально возможным током и с коротким циклом вновь нагревается и становиться проводящим. Однако, по мере остывания всех ИП могут одновременно переключиться два и более ИП и тогда начнется обратный процесс — расфазировки циклических нагревов ИП. На рис. 1. отчетливо видно, что после спонтанного импульса еще достаточно продолжительно происходят всплески сигнала, в соответствии с развиваемой моделью частичная фазировка групп ИП продолжает релаксировать довольно продолжительное время. Заметим, что время восстановления ИП связано не только с локальной температурой, но и с наличием свободного радикала или молекулы H-Cl. Таким образом, цикл восстановления диэлектрических свойств ИП в отсутствие молекулы в прилежащем свободном объеме может затянуться на неопределенный срок, что также наблюдалось в наших экспериментах. Наконец, при увеличении толщины образца пленки композита в несколько раз существенно снижают возможность резких переключений, что в рамках данной модели также легко объяснимо: чем больше число элементов, которые должны синхронизовать свои тепловые циклы, тем реже и маловероятней это событие, тем более, что и взаимное влияние на синхронизацию элементов также сокращается.

Рассматриваемая модель имеет достаточно много свободных параметров, в частности, есть максимальное и минимальное значение сопротивление ИП, здесь опираясь на эксперимент можно судить о $R_i$ max, в то время как для $R_i$ min обсуждается, как правило, при самых уникально малые значения, т. е. переходы сопротивления также наблюдаются и в меньших пределах, но они достаточно неустойчивые и никак не фиксированы в плане воспроизводимости. Это остаточное значение сопротивления образца очень мало влияет на динамику термического цикла ИП и, в частности, может составлять как единицы Ом, так и кило Ом, поскольку и в том и в другом случае оно будет меньше балластного и никак не будет влиять на значение тока в цепи.

В наших экспериментах по измерению температурной зависимости сопротивления образцов пластикатов с различным уровнем проводящих примесей [7] отчетливо зафиксированы перепады с «промежуточной» ступенькой, т. е. не сразу на 11 порядков величины, а по мере увеличения температуры на 7-9 порядков величины.

Рис.5. Зависимость объёмного сопротивления $R_V$ (Ом) от величины обратной температуры [7]. 1-чистая поливинилхлоридная (ПВХ) плёнка
3- плёнка из дегидрохлорирового ПВХ. Время термолиза 240 мин при температуре t=190 C

4-плёнка из дегидрохлорирового ПВХ. Время термолиза 320 мин при температуре t=190C

Все образцы имеют толщину 10-12 мкм. Было проведено численное моделирование температурной зависимости в приближении 3х изолирующих промежутков.

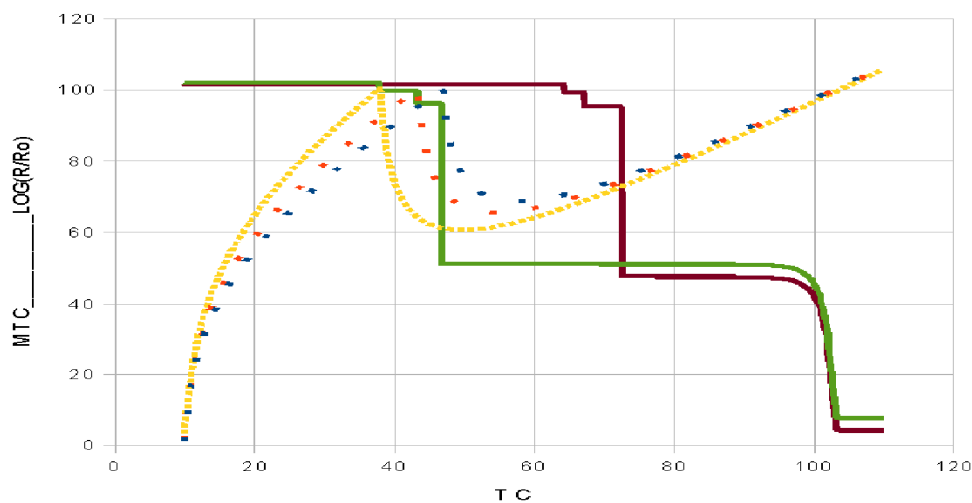

Рис.6 Расчетные зависимости численного моделирования для двух концентраций проводящих примесей в ПВХ –композите в рамках развитого численного моделирования.

Микротемпература изолирующих промежутков T $^0$C при изменении полной температуры образца. Прерывистыми линиями-точками показаны микротемпературы трех изолирующих промежутков для одной концентрации проводящих промежутков в композите ПВХ от полной температуры образца. Легко видеть, что после ступеньки значения микротемпературы начинают падать, однако рост температуры образца как целого не позволяет образцу вернуться в исходное состояние. На рис МТС –микро температура в градусах Цельсия , изменение сопротивления образцов приведено в логарифмическом масштабе.

При высоких концентрациях фрагментов полиацетилена в ПВХ можно наблюдать временный возврат в исходное состояние, см. данные на рис. 7

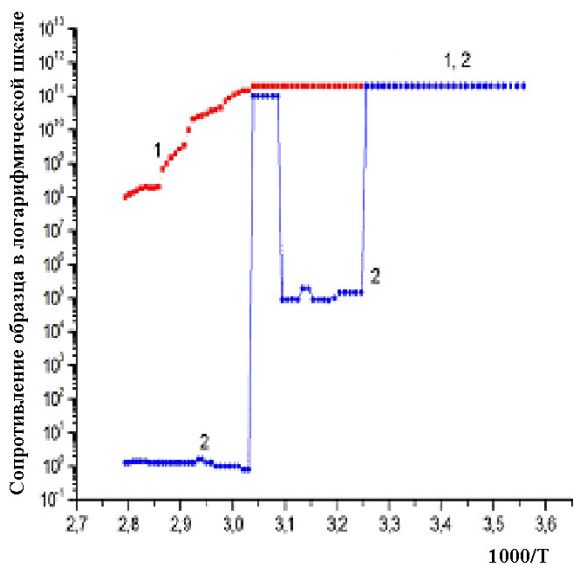

Рис.7 Экспериментальные данные в нормальных координатах Аррениуса: LOG(R/R$_O$ ) от 1000/T . На рис. приведена температурная зависимость для чистого ПВХ и образца ПВХ-композита с максимально высоким содержанием поли ацетилена. На этой экспериментальной кривой, соответствующей композиту (2), виден этап самосинхронизации при достаточно низких температурах, затем начинается остывание и возврат в исходное: фактически полный аналог спонтанного перехода. Затем вследствие дальнейшего разогрева образца вновь происходит самосинхронизации и образец окончательно переключается в проводящее состояние и находится в нем в неограниченно продолжительное время , пока температура близка к 100 $^0$C .

Таким образом, модель позволяет объяснить большинство наблюдаемый спонтанных переходов, а исследования температурной зависимости существенно прибавляет «проверочных» ситуаций, кроме известных спонтанных переходов при комнатной температуре. В частности, при нагревании образца пленки-композита частичная синхронизация реализуется гораздо легче (см рис.3) поскольку появляется больше осциллирующих промежутков, но и достигнуть полной синхронизации становится из-за этого сложнее. Это подтверждается также при измерении температурной зависимости образцов с толщиной 100-250 мкм и более, где можно ожидать (при тех же концентрациях проводящих элементов большего количества изолирующих промежутков) при нагреве увеличивается число «циклирующих» ИП и влиянии нелинейной синхронизации становится слабым. Переключение изолирующих промежутков происходит практически случайно, при приближении температуры к температуре ДГХ.

Выводы.

Развиваемая численная модель позволяет подтвердить сделанные ранее на качественном уровне предположения о тепловом характере скачков проводимости и получить расчетные кривые, похожие практически на все «аномальные» наблюдаемые экспериментально явления переключений проводимости на много порядков величины, в том числе при измерении температурной зависимости проводимости. Сделан собственно первый шаг и намечен путь расчета самосогласованного решения задачи численного моделирования гигантских скачков проводимости. Отдельно стоит отметить простой и бесхитростный механизм, лежащий в основе «спонтанных» само- переключений проводимости. Как следует из самой модели и численного моделирования момент возникновения спонтанного перехода просто связан с самосинхронизацией циклических изменений температуры в нескольких ИП композита в тонком токовом канале. Вероятно развитие этой работы позволит более точно оценить геометрические параметры канала. С практической точки зрения важен тот факт, что в достаточно толстых пленках, содержащих «статистически» большое число ИП - скачки и аномалии практически незаметны по двум причинам, во-первых при переходе в СВП одного из многих ИП полный ток практически не изменяется, следовательно эффект самосинхронизации будет минимален как и собственно изменение полного сопротивления. Таким образом, получаем чисто зависимость по типу Аррениуса, что можно использовать в частности для изготовления полимерного датчика температуры. В заключение автор выражает признательность Л.А.Апресяну за внимательное прочтение статьи и ряд полезных замечаний.

Литература

1. Shklovskii B.I, Éfros A.L. "Percolation theory and conductivity of strongly inhomogeneous media", *Sov. Phys. Usp.* **18** 845–862 (1975);

2. A.N.Lachinov, N.V.Vorob`eva, "Electronics of thin wideband polymer layers"// Phys.Usp., vol.49, pp1223-1238, 2006 .

3. Ralf Strümple, "Polymer composite thermistors for temperature and current sensors", J. Appl. Phys., Vol. 80, No. 11, 1 December 1996

4. D.V.Vlasov, L.A. Apresian, "Cascade Model of Conduction Instability and Giant Fluctuations in Polymer Materials and Nanocomposites". *Amer. J. Mater. Science*, Advances in Materials Science and Applications, vol. 2 Iss. 2, 60-65, 2013

5. D.V.Vlasov, "PVC composite internal process of current-voltage time delay formation and conductivity levels lifetimes origin ", arXiv.:1307.3058

6. I. Shlimak, V. Martchenkov, "Switching phenomena in elastic polymer films", Solid State

Communications,  Volume 107, Issue 9, 24 July 1998, Pages 443–446.

7. Д.В.Власов, В.И.Крыштоб, Т.В.Власова, Л.А.Апресян, С.И.Расмагин, "Исследование влияния температуры на электропроводящие свойства образцов пленок из сополимера ПВХ-ПАц ", arXiv: 1312.7101. а так же: Высокомол. соед. Серия А, 2014, в печати.

8. Крыштоб В.И., Власов Д.В., Миронов В.Ф., Апресян Л.А., Власова Т.В., Расмагин С.И., Кураташвили З.А, Соловский А.А. "Короткое замыкание в электрических кабелях с полимерной изоляцией: новый взгляд на причину его возникновения и пути нетрадиционного решения проблемы". Электротехника №5 (2013) , с.39.

9. C Gau, Cheng-Yung Kuo and H S Ko,  " Electron tunneling in carbon nanotube composites"// Nanotechnology, 20, 395705 (6pp), 2009.